\documentclass{article}
\usepackage{spconf,amsmath,graphicx}

\usepackage{cite}
\usepackage{amssymb,amsfonts}
\usepackage{algorithmic}
\usepackage{textcomp}
\usepackage{xcolor}

\usepackage{times,url,bm}
\usepackage[printonlyused, nolist]{acronym}
\usepackage{pgfplots}
\usepackage{booktabs}
\usepackage{pifont}
\usepackage{algorithm}

\usepackage{tikz}
\usetikzlibrary{plotmarks}
\tikzset{every picture/.style={line width=1.5pt}}

\pgfplotsset{compat=1.11}
\newlength\fwidth

\newcommand{\W}{\mathbf{W}}
\newcommand{\w}{\mathbf{w}}

\newcommand{\y}{\mathbf{y}}
\newcommand{\x}{\mathbf{x}}

\newcommand{\V}{\mathbf{C}}

\newcommand{\FreqIdx}{f}
\newcommand{\FreqIdxMax}{F}
\newcommand{\BlockIdx}{n}
\newcommand{\BlockIdxMax}{N}
\newcommand{\ChannelIdx}{k}
\newcommand{\ChannelIdxMax}{K}

\newcommand{\yFreqVec}{\underline{\y}}

\newcommand{\Allw}{\mathbf{w}}

\newcommand{\transp}{^\text{T}}
\newcommand{\herm}{^\text{H}}
\newcommand{\inv}{^{-1}}
\newcommand{\Expect}[1]{\hat{\mathbb{E}}\left\lbrace #1 \right\rbrace}

\newcommand{\MMstep}[1]{\mathbf{f}\left( #1 \right)}

\newcommand{\FixedPoint}[1]{\Delta\mathbf{f}\left( #1 \right)}
\newcommand{\FixedPointDouble}[1]{\Delta^2\mathbf{f}\left( #1 \right)}

\newcommand{\SQUSec}[1]{\Delta\mathbf{g}\left( #1 \right)}

\newcommand{\Walter}[1]{#1}
\newcommand{\WalterTwo}[1]{#1}

\def\x{{\mathbf x}}

\title{Accelerating Auxiliary Function-based Independent Vector Analysis}
\name{Andreas Brendel and Walter Kellermann\thanks{\WalterTwo{This work was funded by the Deutsche Forschungsgemeinschaft (DFG, German Research Foundation) -- 282835863.}}}
\address{\textit{Multimedia Communications and Signal Processing,}
	\textit{Friedrich-Alexander-Universit\"at Erlangen-N\"urnberg,}\\
	Cauerstr. 7, D-91058 Erlangen, Germany,
	e-mail: \texttt{Andreas.Brendel@FAU.de}}
\begin{document}
	
	\begin{acronym}
		\acro{STFT}{Short-Time Fourier Transform}
		\acro{PSD}{Power Spectral Density}
		\acro{PDF}{Probability Density Function}
		\acro{RIR}{Room Impulse Response}
		\acro{FIR}{Finite Impulse Response}
		\acro{FFT}{Fast Fourier Transform}
		\acro{DFT}{Discrete Fourier Transform}
		\acro{ICA}{Independent Component Analysis}
		\acro{IVA}{Independent Vector Analysis}
		\acro{TRINICON}{TRIple-N Independent component analysis for CONvolutive mixtures}
		\acro{FD-ICA}{Frequency Domain ICA}
		\acro{BSS}{Blind Source Separation}
		\acro{NMF}{Nonnegative Matrix Factorization}
		\acro{MM}{Majorize-Minimize}
		\acro{MAP}{Maximum A Posteriori}
		\acro{RTF}{Relative Transfer Function}
		\acro{AuxIVA}{Auxiliary Function IVA}
		\acro{FD-ICA}{Frequency-Domain Independent Component Analysis}
		\acro{DOA}{Direction of Arrival}
		\acro{SNR}{Signal-to-Noise Ratio}
		\acro{SIR}{Signal-to-Interference Ratio}
		\acro{SDR}{Signal-to-Distortion Ratio}
		\acro{SAR}{Signal-to-Artefact Ratio}
		\acro{GC}{Geometric Constraint}
		\acro{DRR}{Direct-to-Reverberant energy Ratio}
		\acro{ILRMA}{Independent Low Rank Matrix Analysis}
		\acro{IVE}{Independent Vector Extraction}
		\acro{GC-IVA}{Geometric Constraint IVA}
		\acro{SOI}{Sources Of Interest}
		\acro{BG}{Background}
		\acro{MNMF}{Multichannel NMF}
		\acro{IP}{Iterative Projection}
		\acro{EVD}{Eigenvalue Decomposition}
		\acro{GEVD}{Generalized Eigenvalue Decomposition}
		\acro{SQUAREM}{Squared Iterative Methods}
		\acro{EM}{Expectation Maximization}
	\end{acronym}
	
\ninept
\maketitle
\begin{abstract}
Independent Vector Analysis (IVA) is an effective approach for Blind Source Separation (BSS) of convolutive mixtures of audio signals. As a practical realization of an IVA-based BSS algorithm, the so-called AuxIVA update rules based on the Majorize-Minimize (MM) principle have been proposed which allow for fast and computationally efficient optimization of the IVA cost function. \Walter{For many real-time applications, however, update rules for IVA exhibiting even faster convergence are highly desirable.} \Walter{To this end}, we investigate techniques which accelerate the convergence of the AuxIVA update rules \Walter{without \WalterTwo{extra} computational cost}. The efficacy of the proposed methods is \Walter{verified} in experiments \Walter{representing real-world acoustic scenarios}.
\end{abstract}
\section{Introduction}
\label{sec:intro}
In daily-life situations, acoustic sources are usually observed as a mixture, e.g., multiple simultaneously active speakers in the \Walter{much-quoted} cocktail party scenario or a desired acoustic source mixed with \Walter{interferers and} background noise such as, e.g., street noise. \ac{BSS} \cite{vincent_audio_2018,makino_blind_2007} methods aim at separating such mixtures while using only very little information about the \Walter{given scenario}. As typical acoustic scenes within enclosures \Walter{involve} multipath propagation, \ac{ICA}-based approaches relying on instantaneous demixing models \cite{hyvarinen_independent_2001} have been \Walter{extended} to demixing models \Walter{that represent a circular convolution} by solving the \Walter{instantaneous} \ac{BSS} problem in \Walter{individual \ac{DFT} bins} \cite{smaragdis_blind_1998}. However, \Walter{the performance of such narrow-band methods strongly relies on effective solutions for} the well-known internal permutation ambiguity \cite{sawada_robust_2004}. As \Walter{a state-of-the-art method to cope with the internal permutation problem}, \ac{IVA} which uses a multivariate \ac{PDF} as a source model for jointly describing all \ac{DFT} bins has been proposed \cite{kim_blind_2007}.

Real-time applicability of \ac{IVA} calls for fast and efficient optimization \Walter{and} a large variety of methods has been developed since \ac{IVA} has been \Walter{proposed:} Starting with simple gradient and natural gradient algorithms \cite{kim_blind_2007}, step size control mechanisms have been considered to obtain fast and stable convergence \cite{yanfeng_liang_adaptive_2011}. A fast \Walter{fixed-point} algorithm, following the ideas of \Walter{FastICA} \cite{hyvarinen_independent_2001} has been proposed in \cite{lee_fast_2007}. An \ac{EM}-based optimization scheme has been proposed for \ac{IVA} considering additive noise \cite{hao_independent_2010}. Based on the \ac{MM} principle \cite{hunter_tutorial_2004}, fast and stable update rules have been proposed using \Walter{the iterative projection principle} under the name \ac{AuxIVA} \cite{ono_stable_2011}, which do not require tuning parameters such as a step size. \Walter{The latter} can be considered as the gold standard for optimizing the \ac{IVA} cost function. For the special case of two sources and two microphones, even faster update rules based on a generalized eigenvalue decomposition have been developed \cite{ono_fast_2012}.

In this paper, we investigate \Walter{three} methods for \Walter{further} acceleration of the \ac{AuxIVA} update rules. The first method considered here is a Quasi-Newton scheme \cite{nocedal_numerical_2006}, which approximates the differential of the \ac{AuxIVA} update rules \Walter{using} previous \ac{MM} iterates \cite{zhou_quasi-newton_2011}. The second approach uses a gradient-type scheme also called Overrelaxed Bound Optimization \cite{salakhutdinov_adaptive_2003}, which is motivated by the intuition that extending the update of the algorithm into the direction of the current \ac{MM} update may provide accelerated convergence \cite{lange_gradient_1995}. As a third approach, we use the \ac{SQUAREM} technique \cite{varadhan_squared_2004,varadhan_simple_2008}, which has been developed for the acceleration of \ac{EM} algorithms and is based on ideas of extrapolation for increasing the convergence speed of sequences \cite{brezinski_extrapolation_1991}. \Walter{All} investigated acceleration methods are shown to provide faster convergence in experiments with measured \acp{RIR} than the \Walter{original} \ac{AuxIVA} update rules at the same computational cost.

\section{Independent Vector Analysis}
\label{sec:IVA}
In the following, we consider an array of $\ChannelIdxMax$ microphones recording the convolutive mixture of $\ChannelIdxMax$ acoustic sources, i.e., a determined scenario. Using the observed microphone signals \Walter{in the \ac{STFT} domain with} frequency bin $\FreqIdx\in\{1,\dots,\FreqIdxMax\}$ and \Walter{time} frame index $\BlockIdx\in\{1,\dots,\BlockIdxMax\}$
\begin{equation}
	\x_{\FreqIdx,\BlockIdx} = \left[x_{1,\FreqIdx,\BlockIdx},\dots,x_{\ChannelIdxMax,\FreqIdx,\BlockIdx}\right]\transp \in \mathbb{C}^{\ChannelIdxMax}
\end{equation}
the \WalterTwo{demixed signals $\y_{\FreqIdx,\BlockIdx} \in \mathbb{C}^{\ChannelIdxMax}$ are obtained} according to
\begin{equation}
	\y_{\FreqIdx,\BlockIdx} \Walter{= \left[y_{1,\FreqIdx,\BlockIdx},\dots,y_{\ChannelIdxMax,\FreqIdx,\BlockIdx}\right]\transp} = \W_\FreqIdx \x_{\FreqIdx,\BlockIdx},
	\label{eq:demixing_equation}
\end{equation}
\WalterTwo{by the demixing matrix}
\begin{equation}
	\W_\FreqIdx = \begin{bmatrix}
		\w_{1,\FreqIdx},\dots,\w_{\ChannelIdxMax,\FreqIdx}
	\end{bmatrix}\herm \in \mathbb{C}^{\ChannelIdxMax\times\ChannelIdxMax},
\end{equation}
\WalterTwo{with $\w_{\ChannelIdx,\FreqIdx}$ capturing the weights of the $\ChannelIdxMax$-channel MISO system producing the $\FreqIdx$-th \ac{DFT} bin of the $\ChannelIdx$-th demixed signal.}
For notational convenience, we introduce also the broadband demixed signal vector of output channel $\ChannelIdx$
\begin{equation}
	\yFreqVec_{\ChannelIdx,\BlockIdx} = \left[y_{\ChannelIdx,1,\BlockIdx},\dots,y_{\ChannelIdx,\FreqIdxMax,\BlockIdx}\right]\transp  \in \mathbb{C}^{\FreqIdxMax}.
\end{equation}
\Walter{Using} a broadband source model $G(\yFreqVec_{\ChannelIdx,\BlockIdx}) = -\log p(\yFreqVec_{\ChannelIdx,\BlockIdx})$, \Walter{where $p(\cdot)$ is the multivariate \ac{PDF} capturing all complex-valued \ac{STFT} bins of the $\ChannelIdx$th output channel at time frame $\BlockIdx$,} \ac{IVA} aims at separating the sources using the demixing matrices $\W_\FreqIdx$ of all frequency bins determined by \Walter{minimizing the cost function}
\begin{equation}
	J(\Allw) = \sum_{\ChannelIdx=1}^{\ChannelIdxMax} \Expect{G\left(\yFreqVec_{\ChannelIdx,\BlockIdx}\right)} - 2\sum_{\FreqIdx=1}^{\FreqIdxMax}\log \left\vert \det \W_\FreqIdx\right\vert,
	\label{eq:IVA_cost_function}
\end{equation}
where $\Expect{\cdot} = \frac{1}{\BlockIdxMax} \sum_{\BlockIdx=1}^{\BlockIdxMax}(\cdot)$ denotes the averaging operator and 
\begin{equation}
	\Allw = \begin{bmatrix}
		\w_{1,1}\transp,\dots,\w_{\ChannelIdxMax,\FreqIdxMax}\transp
	\end{bmatrix}\transp \in \mathbb{C}^{\ChannelIdxMax\FreqIdxMax}
	\label{eq:def_W_overall}
\end{equation}
the concatenation of demixing vectors of all channels and frequency bins.
For \Walter{minimizing} the cost function \eqref{eq:IVA_cost_function}, the \ac{MM} principle is used in \cite{ono_stable_2011}. Hereby, an upper bound $Q$ for the cost function $J$ is constructed which is easier to optimize and fulfills the properties of majorization and tangency, i.e.,
\begin{equation}
	J(\Allw) \leq Q(\Allw\vert\Allw^{(l)}) \ \text{and} \ J(\Allw^{(l)}) = Q(\Allw^{(l)}\vert\Allw^{(l)}),
	\label{eq:MM_properties}
\end{equation}
where $\Allw^{(l)}$ denotes the concatenation of all demixing vectors \eqref{eq:def_W_overall} determined in iteration \mbox{$l\in \{1,\dots,L\}$}.

The \ac{MM} algorithm iterates between two steps: construction of the upper bound $Q(\Allw\vert\Allw^{(l)})$ by the recent update $\Allw^{(l)}$ to ensure \eqref{eq:MM_properties} and optimization of this upper bound to obtain $\Allw^{(l+1)}$. To construct the upper bound for supergaussian source models $G(\cdot)$ the following inequality has been proposed \cite{ono_stable_2011} \vspace{-2pt}
\begin{equation}
	\Expect{G\left(\yFreqVec_{\ChannelIdx,\BlockIdx}\right)} \leq \frac{1}{2}\sum_{\FreqIdx=1}^{\FreqIdxMax}\left(\w_{\ChannelIdx,\FreqIdx}\herm \V_\FreqIdx^{\ChannelIdx,(l)}\w_{\ChannelIdx,\FreqIdx}\right) +\text{const.}\vspace{-1pt}
	\label{eq:ono_inequality}
\end{equation}
Hereby, $\V_\FreqIdx^{\ChannelIdx,(l)}$ denotes a covariance matrix of the \Walter{observed} signals\vspace{-2pt}
\begin{equation}
	\V_\FreqIdx^{\ChannelIdx,(l)} = \Expect{\frac{G'(r_{\ChannelIdx,\BlockIdx}^{(l)})}{r_{\ChannelIdx,\BlockIdx}^{(l)}}\x_{\FreqIdx,\BlockIdx}\x_{\FreqIdx,\BlockIdx}\herm},\vspace{-1pt}
	\label{eq:weighted_covMat}
\end{equation}
weighted by a factor dependent on the short-time broadband signal magnitude of source $\ChannelIdx$
\begin{equation}
	r_{\ChannelIdx,\BlockIdx}^{(l)} = \left\Vert \yFreqVec_{\ChannelIdx,\BlockIdx}^{(l)} \right\Vert_2 = \sqrt{\sum_{\FreqIdx = 1}^{\FreqIdxMax} \left\vert \left(\w_{\ChannelIdx,\FreqIdx}^{(l)}\right)\herm\x_{\FreqIdx,\BlockIdx} \right\vert^2}.
	\label{eq:signal_energy}
\end{equation}
Application of inequality \eqref{eq:ono_inequality} to the cost function \eqref{eq:IVA_cost_function} yields the upper bound $Q$, which can be \Walter{minimized using} the iterative projection technique \cite{ono_stable_2011} \Walter{stipulating} the following \Walter{update}
\begin{equation}
\Walter{\w_{\ChannelIdx,\FreqIdx}^{(l+1)} = \frac{\left(\W_\FreqIdx^{(l)}\V_\FreqIdx^{\ChannelIdx,(l)}\right)^{-1}\mathbf{e}_\ChannelIdx}{\sqrt{\left(\mathbf{e}\transp_\ChannelIdx\W_\FreqIdx^{(l)}\V_\FreqIdx^{\ChannelIdx,(l)}\left(\W_\FreqIdx^{(l)}\right)\herm\right)^{-1}\mathbf{e}_\ChannelIdx}},}
\label{eq:normalization}
\end{equation}
where $\mathbf{e}_\ChannelIdx$ is the canonical basis vector with a one at the $\ChannelIdx$th position.
\Walter{A complete iteration  for the AuxIVA update} is summarized in Alg.~\ref{alg:pseudocode_IVA}. 
\begin{algorithm}
	\caption{AuxIVA: $\Allw^{(l+1)} = \MMstep{\Allw^{(l)}}$}
	\label{alg:pseudocode_IVA}
	\begin{algorithmic}
		\STATE \textbf{INPUT:} $\Allw^{(l)}$
		\FOR{$\ChannelIdx=1$ \TO $\ChannelIdxMax$}\vspace{5pt}
		\STATE $r_{\ChannelIdx,\BlockIdx}^{(l)} = \sqrt{\sum_{\FreqIdx = 1}^{\FreqIdxMax} \vert (\w_{\ChannelIdx,\FreqIdx}^{(l)})\herm\x_{\FreqIdx,\BlockIdx}\vert^2}$ $\forall \BlockIdx$\vspace{5pt} 
		\FOR{$\FreqIdx=1$ \TO $\FreqIdxMax$}\vspace{2pt}
		\STATE $\V_\FreqIdx^{\ChannelIdx,(l)} = \Expect{\frac{G'(r_{\ChannelIdx,\BlockIdx}^{(l)})}{r_{\ChannelIdx,\BlockIdx}^{(l)}}\x_{\FreqIdx,\BlockIdx}\x_{\FreqIdx,\BlockIdx}\herm}$\vspace{3pt} 
		\STATE \Walter{$\w_{\ChannelIdx,\FreqIdx}^{(l+1)} = \frac{\left(\W_\FreqIdx^{(l)}\V_\FreqIdx^{\ChannelIdx,(l)}\right)^{-1}\mathbf{e}_\ChannelIdx}{\sqrt{\left(\mathbf{e}\transp_\ChannelIdx\W_\FreqIdx^{(l)}\V_\FreqIdx^{\ChannelIdx,(l)}\left(\W_\FreqIdx^{(l)}\right)\herm\right)^{-1}\mathbf{e}_\ChannelIdx}}$}
		\ENDFOR
		\ENDFOR
		\STATE \textbf{OUTPUT:} $\Allw^{(l+1)}$
	\end{algorithmic}
\end{algorithm}
\section{Acceleration Schemes}
\label{sec:acceleration}
In the following, we present three \Walter{methods} for accelerating the convergence of AuxIVA. For convenience, we denote one \Walter{\ac{MM} map in accordance with} Alg.~\ref{alg:pseudocode_IVA} by $\Allw^{(l+1)} = \mathbf{f}(\Allw^{(l)})$. 

After convergence, the \ac{MM} algorithm attains a fixed point
\begin{equation}
	\MMstep{\Allw^{(\infty)}} = \Allw^{(\infty)}.
\end{equation}
Hence, determining this final value $\Allw^{(\infty)}$ corresponds to finding a root of
\begin{equation}
	\FixedPoint{\Allw} = \MMstep{\Allw}- \Allw = \mathbf{0}_{\ChannelIdxMax\FreqIdxMax\times1}.
	\label{eq:fixedPoint_root}
\end{equation}
This problem can be solved by Newton's method \cite{zhou_quasi-newton_2011}
\begin{align}
	\Allw^{(l+1)} &= \Allw^{(l)} - \mathrm{d}\FixedPoint{\Allw^{(l)}}\inv \FixedPoint{\Allw^{(l)}}\label{eq:Newton}
\end{align}
where the differential of $\Delta\mathbf{f}(\Allw^{(l)})$ is denoted by $\mathrm{d}\Delta\mathbf{f}(\Allw^{(l)}) = \mathrm{d}\mathbf{f}(\Allw^{(l)})-\mathbf{I}_{\ChannelIdxMax\FreqIdxMax}$. \Walter{In the following, we present three acceleration methods which can be derived from the Newton-type update \eqref{eq:Newton}.}
\subsection{Quasi-Newton}
\label{sec:newton}
As a first acceleration scheme, we \Walter{apply} the Quasi-Newton approximation of \WalterTwo{(see, e.g., \cite{zhou_quasi-newton_2011}) to \eqref{eq:Newton}}. Here, the differential of the \ac{MM} map $\mathrm{d}\mathbf{f}(\Allw^{(l)})$ is approximated by a matrix $\mathbf{M}$
\begin{equation}
	\mathrm{d}\MMstep{\Allw^{(l)}} \approx \mathbf{M}\in \mathbb{C}^{\ChannelIdxMax\FreqIdxMax\times\ChannelIdxMax\FreqIdxMax},
\end{equation}
which is constructed by so-called secant approximations \cite{nocedal_numerical_2006}
\begin{equation}
	\mathbf{M}\FixedPoint{\Allw^{(l)}} = \FixedPointDouble{\Allw^{(l)}}.
\end{equation}
Hereby, we introduced the following abbreviation
\begin{equation}
	\FixedPointDouble{\Allw^{(l)}} = \mathbf{f}\circ \MMstep{\Allw^{(l)}} - \MMstep{\Allw^{(l)}}
\end{equation}
and $(\cdot)\circ(\cdot)$ denotes the concatenation of functions. Multiple secant approximations\Walter{, we denote their number by $q$,} have to be chosen in order to obtain decent results. \Walter{This can be conveniently expressed in} matrix notation as
\begin{equation}
\mathbf{M}\mathbf{U} = \mathbf{V} \Walter{\quad \mathrm{where} \quad\mathbf{U}, \mathbf{V}\in\mathbb{C}^{\ChannelIdxMax\FreqIdxMax\times q}},
\label{eq:secantApprox_matrix}
\end{equation}
\Walter{i.e., we would obtain, e.g., $\mathbf{U} = [\mathbf{f}(\Allw^{(l)}),\mathbf{f}(\Allw^{(l-1)})]$ for $q=2$.} As a solution for $\mathbf{M}$ which minimizes its Frobenius norm and obeys \eqref{eq:secantApprox_matrix}, the following expression has been derived \cite{zhou_quasi-newton_2011}
\begin{equation}
	\mathbf{M} = \mathbf{V}\left(\mathbf{U}\herm\mathbf{U}\right)\inv\mathbf{U}\herm.
\end{equation} 
Insertion into \eqref{eq:Newton} and application of the matrix inversion lemma yields \cite{zhou_quasi-newton_2011}
\begin{equation}
	\Allw^{(l+1)} = \MMstep{\Allw^{(l)}} - \mathbf{V}\left[\mathbf{U}\herm\mathbf{U} - \mathbf{U}\herm\mathbf{V}\right]\inv \mathbf{U}\herm \FixedPoint{\Allw^{(l)}}.
\end{equation}
Note that the matrix to be inverted here is of dimension $q\times q$, i.e., small \Walter{relative to} the length of $\Allw$, and hence the inversion is computationally cheap. One update of the Quasi-Newton algorithm is summarized in Alg.~\ref{alg:pseudocode_newton}.

\begin{algorithm}
	\caption{Quasi-Newton}
	\label{alg:pseudocode_newton}
	\begin{algorithmic}
		\STATE \textbf{INPUT:} $\Allw^{(l)}$\vspace{2pt}
		\STATE $\FixedPoint{\Allw^{(l)}} = \MMstep{\Allw^{(l)}} - \Allw^{(l)}$\vspace{2pt}
		\STATE Construct $\mathbf{V}$ and $\mathbf{U}$\vspace{2pt}
		\STATE $\Allw^{(l+1)} = \MMstep{\Allw^{(l)}} - \mathbf{V}\left[\mathbf{U}\herm\mathbf{U} - \mathbf{U}\herm\mathbf{V}\right]\inv \mathbf{U}\herm \FixedPoint{\Allw^{(l)}}$\vspace{2pt}
		\STATE \textbf{OUTPUT:} $\Allw^{(l+1)}$
	\end{algorithmic}
\end{algorithm}

\subsection{Gradient Approximation}
\label{sec:momentum}
By approximating the differential of \eqref{eq:fixedPoint_root} by a scaled identity matrix
\begin{equation}
	\mathrm{d}\FixedPoint{\Allw^{(l)}} \approx \frac{1}{\mu}\mathbf{I}_{\ChannelIdxMax\FreqIdxMax}
\end{equation}
we obtain \Walter{with \eqref{eq:Newton}} a gradient-type algorithm with step size \mbox{$\mu \leq -1$}
\begin{equation}
	\Allw^{(l+1)} = \Allw^{(l)} - \mu\FixedPoint{\Allw^{(l)}},
	\label{eq:steepestDescent_Update}
\end{equation}
which operates on the results of the \ac{MM} iterations. Note that a step size of $\mu = -1$ corresponds to the original \ac{MM} algorithm and values above $-1$ will slow down convergence.
There are many options for the choice of $\mu$ (see, e.g., \cite{varadhan_simple_2008}), where line search methods \cite{nocedal_numerical_2006} would be a natural choice. However, \Walter{the calculation of an adaptive step size adds significant computational load to the algorithm, e.g., caused by the evaluation of the cost function \eqref{eq:IVA_cost_function} for line search approaches}. Hence, we will use a fixed step size here.
\subsection{SQUAREM}
\label{sec:squarem}
In the following, we review the \ac{SQUAREM} method, which has been introduced and extensively used for the acceleration of \ac{EM} algorithms \cite{varadhan_squared_2004,varadhan_simple_2008}. Let denote $\mathbf{z}^{(l)}$ the outcome of one gradient update according to \eqref{eq:steepestDescent_Update} with step size~$\alpha$
\begin{equation}
	\mathbf{z}^{(l)} = \Allw^{(l)} - \alpha \FixedPoint{\Allw^{(l)}}.
\end{equation}
The main idea of \ac{SQUAREM} is to square this update, i.e., to subsequently perform another gradient update to obtain the next iterate
\begin{align}
	&\Allw^{(l+1)} = \mathbf{z}^{(l)} - \alpha \FixedPoint{\mathbf{z}^{(l)}}\\
	&\ \ = \Allw^{(l)} - \alpha \FixedPoint{\Allw^{(l)}} -\alpha \left[\left(\MMstep{\Allw^{(l)}} - \alpha \FixedPointDouble{\Allw^{(l)}}\right) \right.  \dots \nonumber\\
	&\quad \dots \left. -\left(\Allw^{(l)} - \alpha \FixedPoint{\Allw^{(l)}}\right)\right]\nonumber\\
	&\ \  =\Allw^{(l)} - 2\alpha \FixedPoint{\Allw^{(l)}} + \alpha^2 \SQUSec{\Allw^{(l)}},
\end{align}
where we introduced the term
\begin{equation}
	\SQUSec{\Allw^{(l)}} = \FixedPointDouble{\Allw^{(l)}} - \FixedPoint{\Allw^{(l)}}.
\end{equation}
One iteration of the \ac{SQUAREM} algorithm is summarized in Alg.~\ref{alg:pseudocode_SQUAREM}.
\begin{algorithm}
	\caption{SQUAREM}
	\label{alg:pseudocode_SQUAREM}
	\begin{algorithmic}
		\STATE \textbf{INPUT:} $\Allw^{(l)}$\vspace{2pt}
		\STATE $\FixedPoint{\Allw^{(l)}} = \MMstep{\Allw^{(l)}} - \Allw^{(l)}$\vspace{2pt}
		\STATE $\SQUSec{\Allw^{(l)}} = \FixedPointDouble{\Allw^{(l)}} - \FixedPoint{\Allw^{(l)}}$\vspace{2pt}
		\STATE $\alpha = -\frac{\Vert \SQUSec{\Allw^{(l)}}\Vert_2}{\Vert \FixedPoint{\Allw^{(l)}}\Vert_2}$\vspace{2pt}
		\STATE $\Allw^{(l+1)} = \Allw^{(l)} - \alpha \FixedPoint{\Allw^{(l)}} + \alpha^2 \SQUSec{\Allw^{(l)}}$\vspace{2pt}
		\STATE \textbf{OUTPUT:} $\Allw^{(l+1)}$
	\end{algorithmic}
\end{algorithm}
\section{Experiments}
\label{sec:experiments}
In the following, we discuss the practical realization of the acceleration methods introduced above and present experimental results.
For the Quasi-Newton method, we constructed the matrices $\mathbf{U}$ and $\mathbf{V}$ representing the secant constraints by using three values for $\Delta\mathbf{f}(\Allw^{(l)})$ and two of $\Delta^2\mathbf{f}(\Allw^{(l)})$ \Walter{prior} to the current iteration, i.e., we computed only one \ac{MM} update in each iteration. Using two \ac{MM} updates per iteration as suggested in \cite{zhou_quasi-newton_2011} did not \Walter{yield better} results in our experiments.

The step size $\mu$ of the gradient algorithm is chosen to be constant for simplicity. For the choice of a step size, convergence speed has to be traded off against stability. Here, a value of $\mu = -1.8$ showed good results in our experiments.
The step size $\alpha$ for the \ac{SQUAREM} algorithm is chosen to be \cite{varadhan_simple_2008}
\begin{equation}
	\alpha = -\frac{\Vert \SQUSec{\Allw^{(l)}}\Vert_2}{\Vert \FixedPoint{\Allw^{(l)}}\Vert_2},
\end{equation}
which is a quite common choice for the \ac{SQUAREM} algorithm \cite{zhao_unified_2017}. This expression for the step size compares the relative change in $\Allw$ by applying the \ac{MM} map once with the corresponding change by applying it twice and weight the first-order $\Delta\mathbf{f}(\Allw^{(l)})$ and second-order update $\Delta\mathbf{g}(\Allw^{(l)})$ accordingly.
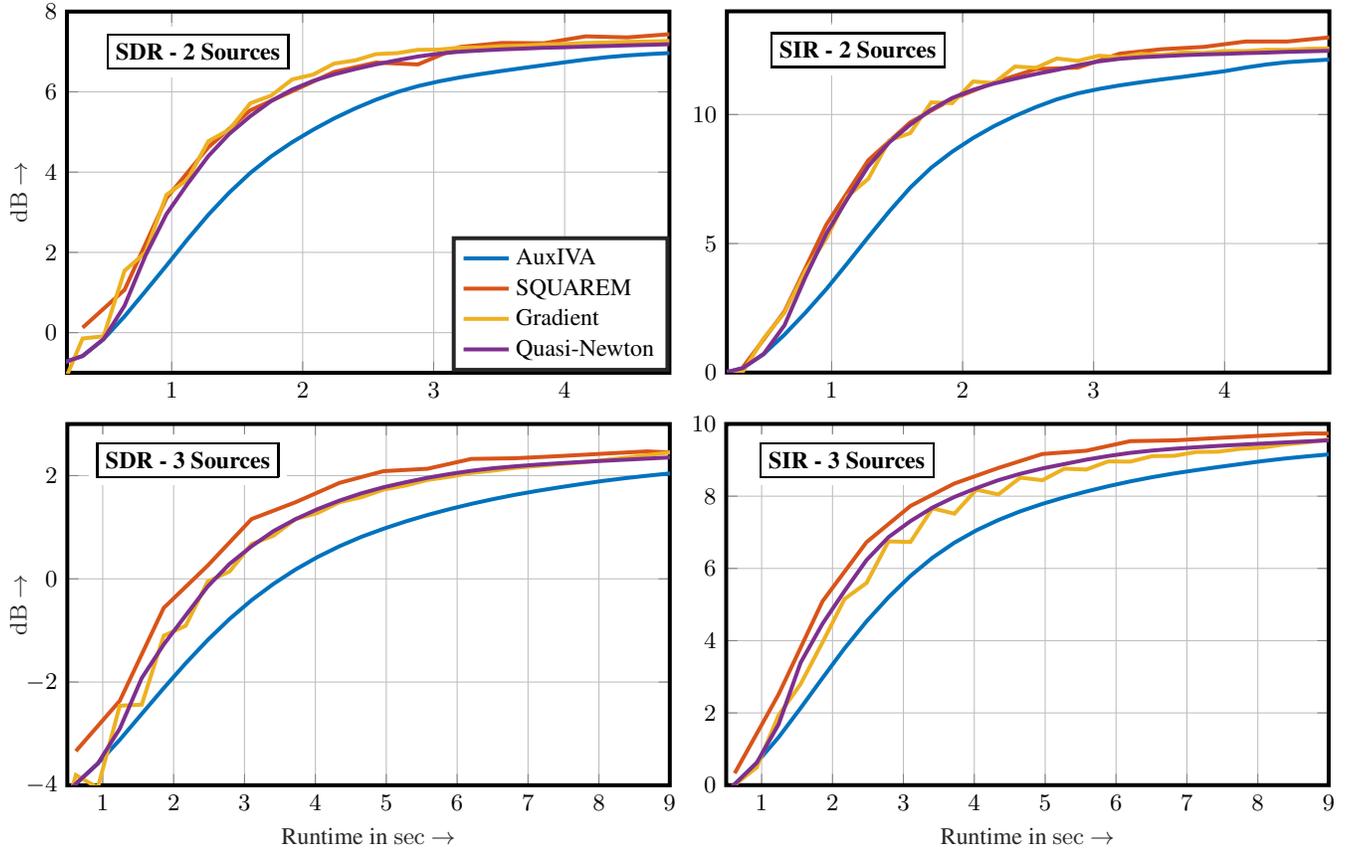
\begin{figure*}[h!]
	\setlength\fwidth{0.45\textwidth}
%
%
\definecolor{mycolor1}{rgb}{0.00000,0.44700,0.74100}%
\definecolor{mycolor2}{rgb}{0.85000,0.32500,0.09800}%
\definecolor{mycolor3}{rgb}{0.92900,0.69400,0.12500}%
\definecolor{mycolor4}{rgb}{0.49400,0.18400,0.55600}%
\begin{tikzpicture}

\begin{axis}[%
width=1\fwidth,
height=0.6\fwidth,
at={(0\fwidth,0\fwidth)},
scale only axis,
xmin=0.2,
xmax=4.8,
xlabel style={font=\color{white!15!black}},
ymin=-1,
ymax=8,
ylabel style={font=\color{white!15!black}},
ylabel={$\mathrm{dB}$ $\rightarrow$},
axis background/.style={fill=white},
title style={font=\bfseries},
xmajorgrids,
ymajorgrids,
legend style={at = {(1,0.378)}, legend cell align=left, align=left, draw=white!15!black}
]

\node[fill = white] at (axis cs: 1.2,7) {\boxed{\textbf{SDR - 2 Sources}}};

\addplot [color=mycolor1]
  table[row sep=crcr]{%
0.16	-0.7602\\
0.32	-0.5807\\
0.48	-0.161\\
0.64	0.4055\\
0.8	1.034\\
0.96	1.671\\
1.12	2.322\\
1.28	2.939\\
1.44	3.492\\
1.6	3.98\\
1.76	4.394\\
1.92	4.746\\
2.08	5.056\\
2.24	5.338\\
2.4	5.586\\
2.56	5.805\\
2.72	5.993\\
2.88	6.143\\
3.04	6.261\\
3.2	6.359\\
3.36	6.444\\
3.52	6.521\\
3.68	6.594\\
3.84	6.665\\
4	6.736\\
4.16	6.805\\
4.32	6.863\\
4.48	6.907\\
4.64	6.94\\
4.8	6.967\\
};
\addlegendentry{AuxIVA}

\addplot [color=mycolor2]
  table[row sep=crcr]{%
0.32	0.1245\\
0.64	1.068\\
0.96	3.351\\
1.28	4.621\\
1.6	5.539\\
1.92	6.016\\
2.24	6.497\\
2.56	6.73\\
2.88	6.683\\
3.2	7.118\\
3.52	7.218\\
3.84	7.211\\
4.16	7.383\\
4.48	7.357\\
4.8	7.436\\
};
\addlegendentry{SQUAREM}

\addplot [color=mycolor3]
  table[row sep=crcr]{%
0.16	-1.378\\
0.32	-0.1444\\
0.48	-0.09012\\
0.64	1.543\\
0.8	1.999\\
0.96	3.434\\
1.12	3.81\\
1.28	4.77\\
1.44	5.058\\
1.6	5.715\\
1.76	5.907\\
1.92	6.309\\
2.08	6.43\\
2.24	6.704\\
2.4	6.784\\
2.56	6.94\\
2.72	6.964\\
2.88	7.047\\
3.04	7.052\\
3.2	7.101\\
3.36	7.105\\
3.52	7.141\\
3.68	7.152\\
3.84	7.182\\
4	7.192\\
4.16	7.217\\
4.32	7.229\\
4.48	7.247\\
4.64	7.254\\
4.8	7.269\\
};
\addlegendentry{Gradient}

\addplot [color=mycolor4]
  table[row sep=crcr]{%
0.16	-0.7602\\
0.32	-0.5807\\
0.48	-0.161\\
0.64	0.6708\\
0.8	1.922\\
0.96	2.955\\
1.12	3.704\\
1.28	4.4\\
1.44	4.954\\
1.6	5.392\\
1.76	5.772\\
1.92	6.058\\
2.08	6.273\\
2.24	6.436\\
2.4	6.568\\
2.56	6.685\\
2.72	6.787\\
2.88	6.881\\
3.04	6.951\\
3.2	7.001\\
3.36	7.031\\
3.52	7.058\\
3.68	7.077\\
3.84	7.098\\
4	7.11\\
4.16	7.128\\
4.32	7.142\\
4.48	7.158\\
4.64	7.171\\
4.8	7.186\\
};
\addlegendentry{Quasi-Newton}

\end{axis}
\end{tikzpicture}%
\hspace{5pt}
%
%
\definecolor{mycolor1}{rgb}{0.00000,0.44700,0.74100}%
\definecolor{mycolor2}{rgb}{0.85000,0.32500,0.09800}%
\definecolor{mycolor3}{rgb}{0.92900,0.69400,0.12500}%
\definecolor{mycolor4}{rgb}{0.49400,0.18400,0.55600}%
\begin{tikzpicture}

\begin{axis}[%
width=1\fwidth,
height=0.6\fwidth,
at={(0\fwidth,0\fwidth)},
scale only axis,
xmin=0.2,
xmax=4.8,
xlabel style={font=\color{white!15!black}},
ymin=0,
ymax=14,
ylabel style={font=\color{white!15!black}},
axis background/.style={fill=white},
title style={font=\bfseries},
xmajorgrids,
ymajorgrids,
legend style={legend cell align=left, align=left, draw=white!15!black}
]

\node[fill = white] at (axis cs: 1.2,12.5) {\boxed{\textbf{SIR - 2 Sources}}};

\addplot [color=mycolor1]
  table[row sep=crcr]{%
0.16	-0.03048\\
0.32	0.1688\\
0.48	0.7142\\
0.64	1.469\\
0.8	2.32\\
0.96	3.237\\
1.12	4.234\\
1.28	5.254\\
1.44	6.248\\
1.6	7.167\\
1.76	7.939\\
1.92	8.559\\
2.08	9.089\\
2.24	9.551\\
2.4	9.943\\
2.56	10.29\\
2.72	10.59\\
2.88	10.82\\
3.04	10.99\\
3.2	11.13\\
3.36	11.25\\
3.52	11.36\\
3.68	11.46\\
3.84	11.57\\
4	11.68\\
4.16	11.82\\
4.32	11.94\\
4.48	12.03\\
4.64	12.08\\
4.8	12.13\\
};

\addplot [color=mycolor2]
  table[row sep=crcr]{%
0.32	0.1911\\
0.64	2.369\\
0.96	5.737\\
1.28	8.229\\
1.6	9.697\\
1.92	10.6\\
2.24	11.24\\
2.56	11.76\\
2.88	11.82\\
3.2	12.36\\
3.52	12.53\\
3.84	12.62\\
4.16	12.83\\
4.48	12.83\\
4.8	12.99\\
};

\addplot [color=mycolor3]
  table[row sep=crcr]{%
0.16	0.03859\\
0.32	0.02029\\
0.48	1.322\\
0.64	2.32\\
0.8	3.953\\
0.96	5.205\\
1.12	6.827\\
1.28	7.508\\
1.44	9.001\\
1.6	9.278\\
1.76	10.47\\
1.92	10.44\\
2.08	11.28\\
2.24	11.22\\
2.4	11.86\\
2.56	11.81\\
2.72	12.17\\
2.88	12.08\\
3.04	12.28\\
3.2	12.22\\
3.36	12.35\\
3.52	12.31\\
3.68	12.41\\
3.84	12.39\\
4	12.46\\
4.16	12.46\\
4.32	12.51\\
4.48	12.51\\
4.64	12.55\\
4.8	12.56\\
};

\addplot [color=mycolor4]
  table[row sep=crcr]{%
0.16	-0.03048\\
0.32	0.1688\\
0.48	0.7142\\
0.64	1.835\\
0.8	3.698\\
0.96	5.39\\
1.12	6.729\\
1.28	7.996\\
1.44	8.921\\
1.6	9.607\\
1.76	10.19\\
1.92	10.64\\
2.08	10.96\\
2.24	11.2\\
2.4	11.39\\
2.56	11.57\\
2.72	11.73\\
2.88	11.91\\
3.04	12.06\\
3.2	12.16\\
3.36	12.22\\
3.52	12.26\\
3.68	12.3\\
3.84	12.33\\
4	12.35\\
4.16	12.38\\
4.32	12.4\\
4.48	12.43\\
4.64	12.45\\
4.8	12.47\\
};

\end{axis}
\end{tikzpicture}%
\\
%
%
\definecolor{mycolor1}{rgb}{0.00000,0.44700,0.74100}%
\definecolor{mycolor2}{rgb}{0.85000,0.32500,0.09800}%
\definecolor{mycolor3}{rgb}{0.92900,0.69400,0.12500}%
\definecolor{mycolor4}{rgb}{0.49400,0.18400,0.55600}%
\begin{tikzpicture}

\begin{axis}[%
width=1\fwidth,
height=0.6\fwidth,
at={(0\fwidth,0\fwidth)},
scale only axis,
xmin=0.5,
xmax=9,
xlabel style={font=\color{white!15!black}},
xlabel={Runtime in $\mathrm{sec}$ $\rightarrow$},
ymin=-4,
ymax=3,
ylabel style={font=\color{white!15!black}, yshift = -7pt},
ylabel={$\mathrm{dB}$ $\rightarrow$},
axis background/.style={fill=white},
title style={font=\bfseries},
xmajorgrids,
ymajorgrids,
legend style={legend cell align=left, align=left, draw=white!15!black}
]

\node[fill = white] at (axis cs: 2.2,2.3) {\boxed{\textbf{SDR - 3 Sources}}};

\addplot [color=mycolor1]
  table[row sep=crcr]{%
0.31	-4.351\\
0.62	-3.973\\
0.93	-3.58\\
1.24	-3.119\\
1.55	-2.617\\
1.86	-2.118\\
2.17	-1.637\\
2.48	-1.186\\
2.79	-0.7764\\
3.1	-0.413\\
3.41	-0.09613\\
3.72	0.1812\\
4.03	0.4227\\
4.34	0.6324\\
4.65	0.8149\\
4.96	0.9701\\
5.27	1.111\\
5.58	1.237\\
5.89	1.35\\
6.2	1.453\\
6.51	1.546\\
6.82	1.63\\
7.13	1.705\\
7.44	1.773\\
7.75	1.837\\
8.06	1.898\\
8.37	1.951\\
8.68	1.998\\
8.99	2.041\\
9.3	2.081\\
};

\addplot [color=mycolor2]
  table[row sep=crcr]{%
0.62	-3.339\\
1.24	-2.364\\
1.86	-0.564\\
2.48	0.2559\\
3.1	1.159\\
3.72	1.485\\
4.34	1.86\\
4.96	2.087\\
5.58	2.133\\
6.2	2.324\\
6.82	2.337\\
7.44	2.379\\
8.06	2.425\\
8.68	2.47\\
9.3	2.431\\
};

\addplot [color=mycolor3]
  table[row sep=crcr]{%
0.31	-5.364\\
0.62	-3.802\\
0.93	-4.047\\
1.24	-2.459\\
1.55	-2.438\\
1.86	-1.103\\
2.17	-0.9092\\
2.48	-0.05292\\
2.79	0.1421\\
3.1	0.6747\\
3.41	0.8369\\
3.72	1.16\\
4.03	1.275\\
4.34	1.485\\
4.65	1.582\\
4.96	1.729\\
5.27	1.807\\
5.58	1.916\\
5.89	1.977\\
6.2	2.053\\
6.51	2.098\\
6.82	2.152\\
7.13	2.187\\
7.44	2.224\\
7.75	2.256\\
8.06	2.301\\
8.37	2.346\\
8.68	2.397\\
8.99	2.448\\
9.3	2.497\\
};

\addplot [color=mycolor4]
  table[row sep=crcr]{%
0.31	-4.351\\
0.62	-3.973\\
0.93	-3.58\\
1.24	-2.904\\
1.55	-1.923\\
1.86	-1.273\\
2.17	-0.7082\\
2.48	-0.15\\
2.79	0.291\\
3.1	0.6315\\
3.41	0.9267\\
3.72	1.157\\
4.03	1.351\\
4.34	1.517\\
4.65	1.658\\
4.96	1.775\\
5.27	1.871\\
5.58	1.958\\
5.89	2.031\\
6.2	2.094\\
6.51	2.144\\
6.82	2.181\\
7.13	2.213\\
7.44	2.24\\
7.75	2.265\\
8.06	2.287\\
8.37	2.309\\
8.68	2.33\\
8.99	2.352\\
9.3	2.37\\
};

\end{axis}
\end{tikzpicture}%
%
%
\definecolor{mycolor1}{rgb}{0.00000,0.44700,0.74100}%
\definecolor{mycolor2}{rgb}{0.85000,0.32500,0.09800}%
\definecolor{mycolor3}{rgb}{0.92900,0.69400,0.12500}%
\definecolor{mycolor4}{rgb}{0.49400,0.18400,0.55600}%
\begin{tikzpicture}

\begin{axis}[%
width=1\fwidth,
height=0.6\fwidth,
at={(0\fwidth,0\fwidth)},
scale only axis,
xmin=0.5,
xmax=9,
xlabel style={font=\color{white!15!black}},
xlabel={Runtime in $\mathrm{sec}$ $\rightarrow$},
ymin=0,
ymax=10,
ylabel style={font=\color{white!15!black}},
axis background/.style={fill=white},
title style={font=\bfseries},
xmajorgrids,
ymajorgrids,
legend style={legend cell align=left, align=left, draw=white!15!black}
]

\node[fill = white] at (axis cs: 2.2,9) {\boxed{\textbf{SIR - 3 Sources}}};

\addplot [color=mycolor1]
  table[row sep=crcr]{%
0.31	-0.514\\
0.62	0.02481\\
0.93	0.6174\\
1.24	1.327\\
1.55	2.134\\
1.86	2.969\\
2.17	3.787\\
2.48	4.534\\
2.79	5.202\\
3.1	5.793\\
3.41	6.296\\
3.72	6.713\\
4.03	7.055\\
4.34	7.34\\
4.65	7.582\\
4.96	7.786\\
5.27	7.967\\
5.58	8.129\\
5.89	8.276\\
6.2	8.409\\
6.51	8.526\\
6.82	8.63\\
7.13	8.721\\
7.44	8.806\\
7.75	8.888\\
8.06	8.969\\
8.37	9.04\\
8.68	9.1\\
8.99	9.154\\
9.3	9.205\\
};

\addplot [color=mycolor2]
  table[row sep=crcr]{%
0.62	0.3294\\
1.24	2.516\\
1.86	5.093\\
2.48	6.727\\
3.1	7.729\\
3.72	8.351\\
4.34	8.781\\
4.96	9.167\\
5.58	9.258\\
6.2	9.523\\
6.82	9.545\\
7.44	9.612\\
8.06	9.674\\
8.68	9.736\\
9.3	9.735\\
};

\addplot [color=mycolor3]
  table[row sep=crcr]{%
0.31	-0.7181\\
0.62	0.008794\\
0.93	0.4907\\
1.24	1.937\\
1.55	2.798\\
1.86	3.956\\
2.17	5.154\\
2.48	5.597\\
2.79	6.749\\
3.1	6.732\\
3.41	7.664\\
3.72	7.516\\
4.03	8.18\\
4.34	8.048\\
4.65	8.513\\
4.96	8.442\\
5.27	8.761\\
5.58	8.738\\
5.89	8.963\\
6.2	8.956\\
6.51	9.108\\
6.82	9.117\\
7.13	9.223\\
7.44	9.232\\
7.75	9.31\\
8.06	9.346\\
8.37	9.43\\
8.68	9.487\\
8.99	9.572\\
9.3	9.631\\
};

\addplot [color=mycolor4]
  table[row sep=crcr]{%
0.31	-0.514\\
0.62	0.02481\\
0.93	0.6174\\
1.24	1.686\\
1.55	3.39\\
1.86	4.471\\
2.17	5.375\\
2.48	6.231\\
2.79	6.866\\
3.1	7.31\\
3.41	7.684\\
3.72	7.978\\
4.03	8.223\\
4.34	8.443\\
4.65	8.62\\
4.96	8.767\\
5.27	8.889\\
5.58	9.011\\
5.89	9.112\\
6.2	9.199\\
6.51	9.263\\
6.82	9.312\\
7.13	9.354\\
7.44	9.391\\
7.75	9.424\\
8.06	9.456\\
8.37	9.487\\
8.68	9.516\\
8.99	9.544\\
9.3	9.567\\
};

\end{axis}
\end{tikzpicture}%
\vspace{-8pt}
	\caption{Performance of the discussed algorithmic variants in terms of \ac{SDR} and \ac{SIR} w.r.t. runtime of the algorithms \Walter{for a segment of $10\ \mathrm{secs}$ of speech}. 
	The \Walter{plots} are created \Walter{by averaging} results for all three different rooms ($T_{60} = 0.2\,\mathrm{sec},0.4\,\mathrm{sec},0.9\,\mathrm{sec}$) and two different source-array distances ($1\,\mathrm{m}$, $2\,\mathrm{m}$). Each experiment corresponding to a certain room and distance has been repeated 20 times choosing the source signals randomly from a set of four male and four female speech signals. 
The first row of plots shows results for a determined scenario comprising $2$ sources and $2$ microphones, the second row shows results for $3$ sources and $3$ microphones.\vspace{-10pt}
}
	\label{fig:results}
\end{figure*}
For the experimental evaluation, we simulated microphone signals by convolving speech signals randomly chosen from a set of 4 male and 4 female speech signals of about $10\,\mathrm{sec}$ duration with \acp{RIR} measured in three different rooms: two meeting rooms ($T_{60} = 0.2\,\mathrm{s}$ and $T_{60} = 0.4\,\mathrm{s}$) and a seminar room ($T_{60} = 0.9\,\mathrm{s}$). The \acp{RIR} are measured with a linear microphone array with $4.2\,\mathrm{cm}$ spacing between the microphones. Two configurations of \acp{RIR} have been measured in the mentioned enclosures at $1\,\mathrm{m}$ and $2\,\mathrm{m}$ distance from the microphone array: $40^\circ/140^0$ and $40^\circ/90^\circ/140^\circ$ w.r.t. the array axis. As we consider only determined scenarios, the number of sources and microphones was equal in all measurements. White Gaussian noise was added to obtain an \ac{SNR} of $30\,\mathrm{dB}$ at the microphones.

The microphone signals have been transformed into the \ac{STFT} \Walter{domain} by employing a Hamming window of length $2048$ and $50\%$ overlap at a sampling frequency of $16\,\mathrm{kHz}$. The performance of the algorithms has been measured by the \acf{SDR}, \acf{SIR} and \acf{SAR} w.r.t. the unprocessed signals \cite{vincent_performance_2006}. Note that these performance measures are indirect indicators for the convergence of the algorithm, as they do not express the costs to be minimized. However, they can be seen as a strong indicator for the separation quality as experienced by a user.
We used a Laplacian source model, i.e., \mbox{$G(r_{\ChannelIdx,\BlockIdx}(\Allw_\ChannelIdx)) = r_{\ChannelIdx,\BlockIdx}(\Allw_\ChannelIdx)$}, which is a common choice for \ac{IVA} applied to audio signals~\mbox{\cite{kim_blind_2007,ono_stable_2011}}.
The results of the experiments described above are shown in Fig.~\ref{fig:results} in terms of \ac{SDR} and \ac{SIR}. Results for the improvement of the \ac{SAR} are omitted due to space constraints. However, the \ac{SAR} improvement was roughly the same for the investigated methods. Fig.~\ref{fig:results} \Walter{shows} the results for scenarios comprising $2$ sources and $2$ microphones and $3$ sources and $3$ microphones. All three different rooms ($T_{60} = 0.2\,\mathrm{sec},0.4\,\mathrm{sec},0.9\,\mathrm{sec}$) and the two different source-array distances ($1\,\mathrm{m}$, $2\,\mathrm{m}$) have been evaluated by repeating the experiment $20$ times for each configuration, where the source signals are drawn randomly from a set of four male and four female speech signals. \Walter{The mean performance values from these different acoustic conditions \WalterTwo{are} shown for the discussed algorithms over runtime in Fig.~\ref{fig:results}.}

The \ac{SQUAREM}-based method converged after roughly $15$ iterations, all other methods after about $30$ iterations. \Walter{To take into account additional computational cost of more advanced algorithms which increase the convergence rate per iteration} the runtime per iteration has been \Walter{considered in order to obtain a fair comparison}. Here, it turned out that the runtime is dominated by the evaluation of the \ac{MM} map and the additional runtime caused by operations added to the \ac{MM} map was negligible. The runtime per iteration for AuxIVA, the gradient-based and the Quasi-Newton-based method was roughly $0.16\,\mathrm{sec}$ for two sources and $0.27\,\mathrm{sec}$ for three sources on average. Due to the second required \ac{MM} map the \ac{SQUAREM} method needed roughly twice as much runtime per iteration.
These observations have been incorporated into Fig.~\ref{fig:results} by showing the performance of the algorithms in terms of runtime of the algorithm. It can be observed that all algorithms \Walter{converge to similar final values with a slight advantage for the acceleration methods}. However, all acceleration schemes provide significantly faster convergence than \ac{AuxIVA} itself. The gradient-type method and the Quasi-Newton method, both using only a single \ac{MM} map, showed similar convergence speed. The \ac{SQUAREM} method based on two \ac{MM} maps outperforms these methods \Walter{especially for the three-source case} and provides \ac{SDR} and \ac{SIR} improvements in the early convergence phase which are higher by several $\mathrm{dB}$ compared to the \ac{AuxIVA} results at the same runtime requirement.

\section{Conclusions}
\label{sec:conclusions}
We investigated the application of three different schemes for the acceleration of the convergence of the \ac{AuxIVA} update rules. We showed that all three methods increased the convergence speed in terms of \ac{SDR} and \ac{SIR} improvement at the same runtime requirements as \ac{AuxIVA}. The gradient-based approach represents a simple but effective modification of the original algorithm but requires the selection of a suitable step size. In our experiments, a fixed step size showed promising results, but future work should investigate mechanisms to choose this step size automatically. The Quasi-Newton method performed similarly as the gradient-based method and was slightly outperformed by the \ac{SQUAREM} method.

Future work will include \WalterTwo{an} in-depth \WalterTwo{investigation} of \WalterTwo{other} acceleration methods (e.g., \cite{berlinet_parabolic_2009}). Also the application of such acceleration schemes to other \ac{BSS} algorithms, which suffer from slow convergence, e.g., \ac{MNMF} \cite{sawada_multichannel_2013} and \ac{TRINICON} \cite{buchner_generalization_2005}, will be part of future work.
\bibliographystyle{IEEEbib}
\bibliography{literature}

\end{document}